          [2004/06/22 1.2 (KOF); 2001/04/25 1.1 (PWD)]

\documentclass[a4paper,twocolumn]{Gaia2004} 
\usepackage{times}      
\usepackage{epsfig}     
\usepackage{natbib}     
\title{A Prototype for Science Alerts}

\author{N. Wyn Evans,~}
\author{Vasily Belokurov}
\affil{Institute of Astronomy, Madingley Rd, Cambridge, UK}

\bibpunct{(}{)}{;}{a}{}{,}  

\newcommand\bw{{\bf {w}}}
\newcommand\bx{{\bf {x}}}

\begin{document}


\maketitle

\begin{abstract}
Many of the science goals of the GAIA mission, especially for bursting
or time-varying phenomena like supernovae or microlensing, require an
early identification, analysis and release of preliminary data.  The
alerting on rare and unusual events is the scientific equivalent of
the finding of needles in haystacks or the panning for gold-dust in
rivers. In modern signal processing, such novelty detection is
routinely performed with self-organising maps (SOMs), which are an
unsupervised clustering algorithm invented by Kohonen. Here, we
describe the application of SOMS to the classification of data
provided by large-scale surveys such as GAIA and to the despatching of
scientific alerts. We illustrate our ideas by processing the
publically available OGLE II dataset towards the Bulge, identifying
major classes of variable stars (such as novae, small amplitude red
giant variables, eclipsing binaries and so on) and extracting the
rare, discrepant lightcurves from which the alerts can be drawn.
\end{abstract}

\section{Introduction}

For the GAIA mission, our aim is to alert on the most interesting
classes of variable objects so that follow-ups using ground-based
telescopes can begin~\footnote{The web-site of the GAIA science alerts
working group is at ``http://www.ast.cam.ac.uk/$\sim$vasily/sawg''}.  This includes supernovae,
microlensing events, near-Earth asteroids, novae, stars undergoing
rare and interesting phases of evolution (such as helium flash), and
so on (see Belokurov \& Evans 2002, 2003 who provide detailed
simulations). Every class of object will require an individual
trigger.

In the signal processing literature, the technique of self-organising
maps (SOMs) is often used as a mechanism for novelty detection (e.g.,
Markou \& Singh 2003). SOMs are an unsupervised learning or clustering
algorithm invented by Kohonen in 1982. They have already found a
number of applications in astronomy -- although for classification
rather than novelty detection purposes (see Belokurov, Evans \& Feeney
2004).

\begin{figure*}[!ht]
  \begin{center}
\vspace{0.25\hsize}\centering{\tt figure1.gif}\vspace{0.25\hsize}
  \end{center} 
\caption{A SOM constructed with $\sim 10^5$ OGLE II lightcurves with the
nodes colour-coded according to distance. The key is given in the
upper left panel (yellow denoting close and black distant nodes).}
\label{fig:firstsom}
\end{figure*}
\begin{figure*}[!ht]
  \begin{center}
\vspace{0.25\hsize}\centering{\tt figure2.gif}\vspace{0.25\hsize}
  \end{center} 
\caption{The same SOM is shown as in Fig. 1, but each node is now
colour-coded according to the number of hits. The key is given in the
upper left panel (white/yellow denoting populous and red/black
sparsely-populated nodes).}
\label{fig:secondsom}
\end{figure*}
\begin{figure*}[!ht]
  \begin{center}
\vspace{0.25\hsize}\centering{\tt figure3.gif}\vspace{0.25\hsize}
  \end{center} 
\caption{The nodes of the SOM are colour-coded according to distance (as in
Fig. 1) and the key is given in the upper left panel.  Contours of the
bump statistic are overplotted onto the map. The key for the colour of
the contour is given in the upper middle panel (blue denoting a dip
and yellow a high peak). The nodes onto which the known eclipsing
binaries are mapped are marked with a number.  This is the percentile
of the node quantization error distribution corresponding to the
median error. The key for the colour of the number is given in the
upper right panel (black denoting few and white many eclipsing
binaries).}
\label{fig:thirdsom}
\end{figure*}

\section{Ideology}

SOMs are two-dimensional lattices of nodes. Roughly speaking, the
number of nodes of the SOM is the number of distinct classes. This is
always much smaller than the number of different examples in the
dataset. The training algorithm ensures that the nodes represent the
most abundant classes in the dataset. If a pattern is rare, then
necessarily there will be no node allocated near to it.

As a test dataset, let us use lightcurves from the OGLE II (Optical
Gravitational Lensing Experiment) photometric survey.  This is
searching for microlensing events and transits towards the Galactic
bulge.  The second phase of this experiment resulted in a catalogue of
220\,000 $I$ band lightcurves of variable objects constructed using
difference image analysis (Wo\'zniak et al. 2002). Each lightcurve is
replaced by a vector $\bx$ in a large (but finite) dimensional vector
space, called the ``pattern space''.
 
It makes sense to construct the SOMs with high-quality
lightcurves. So, the first job is to select these.  This is done by
allocating a rough measure of signal-to-noise ratio (S/N) to each
lightcurve. The three maximum flux values and three minimum flux
values are used to construct 9 flux differences. In each case, the
noise is computed by adding the flux errors of the individual
measurements in quadrature. This gives 9 estimates of S/N, of which
the minimum is selected to guard against outliers. Only lightcurves
with S/N exceeding 4 and valid $V-I$ colour are selected to give $1.3
\times 10^5$ in total.  Each of these is analysed with a Lomb-Scargle
periodogram. The power spectra are binned in the following way. First,
we identify 6 ranges of interest (corresponding to the period
intervals defined by the endpoints 0.1, 1.1, 3, 9, 30, 100, 1000 in
days). Each range is split into 10 equally-spaced bins in the
frequency domain. The maximum value of the power spectrum in each bin
is found. This gives a crude envelope for the shape of the power
spectrum, which is now scaled so that its maximum value is unity.
This associates each lightcurve with a 60-dimensional vector.  To
this, 5 further pieces of information are added.  The first is a
magnitude difference. From the distribution of flux measurements, the
2nd and 98th percentiles are found and converted to a flux difference
in magnitudes using the zeropoint of the difference image analysis.
The second is the flux difference between the 98th percentile and the
50th (the median). This is normalised by the flux difference between
the 98th and 2nd percentiles to give a number between zero and
unity. This gives us a way of distinguishing between dips and bumps,
and is called the ``bump statistic''. The third is the $V-I$ colour,
while the fourth is the robust kurtosis measure, as defined by Stetson
(1996). The fifth and final input is the difference between the
$\chi^2$ of a constant baseline fit and of a linear fit to the
lightcurve, normalised to unity. So, this input is distributed between
0 (lightcurves with no apparent gradient) and 1 (linear gradient).
Each lightcurve has now been replaced by a 65 dimensional vector in
the pattern space.

\begin{figure*}[!ht]
  \begin{center}
\vspace{0.25\hsize}\centering{\tt figure4.gif}\vspace{0.25\hsize}
  \end{center} \caption{ The nodes of the SOM are coloured according
  to distance (as in Fig. 1) and the key is given in the upper left
  panel.  Contours of the bump statistic are plotted on the map. The
  nodes onto which known cataclysmic variables are mapped are marked
  with a number, using the same convention as in Fig.~3.}
\label{fig:fourthsom}
\end{figure*}
\begin{figure*}[!ht]
  \begin{center}
\vspace{0.25\hsize}\centering{\tt figure5.gif}\vspace{0.25\hsize}
  \end{center}
\caption{
The nodes of the SOM are coloured according to distance (as in Fig. 1)
and the key is given in the upper left panel.  Contours of the
amplitude are plotted on the map.  The key for the colour of the
contour is given in the upper middle panel (blue denoting low amplitude
and yellow high).  The nodes onto which the known small-amplitude red
giant variable stars of type A are mapped are marked with a number,
using the same convention as in Fig.~3}
\label{fig:fifthsom}
\end{figure*}

The standard algorithm for creating a SOM has four steps, namely
(e.g., Haykin 1994, Kohonen 2000):

\noindent
[1] Values for the initial weight vectors $\bw_j (0)$ at each node of
the map are picked, It is simplest of all to pick these starting
conditions randomly.

\noindent
[2] A pattern $\bx$ is chosen from the dataset.

\noindent
[3] The winning node $i(\bx)$ at each iteration $n$ is chosen using a minimum
Euclidean distance criterion
\begin{equation}
i(\bx) = j,\ {\rm corresponding\; to}\ \min || \bx (n) - \bw_j ||
\end{equation}

\noindent
[4] The weights of all the nearby nodes are updated
\begin{equation}
\bw_j (n\!+\!1) = \bw_j (n) + \eta (n) [ \bx(n) - \bw_j (n)], \quad
j \in \Lambda (n)
\end{equation}
where $\eta(n)$ is the learning rate and $\Lambda (n)$ is the
neighbourhood function, which shrinks with iteration number $n$.  Now
return to [2] and perform for the next pattern in the dataset. Once
the whole dataset has been processed, do it again $10^5$ times.

Pictorially, we may imagine a lattice connected with springs. The
springs may be stretched or compressed, but the linkages may not be
broken. The algorithm attempts to match the nodes of the lattice with
the centres of clustering of the dataset in a higher dimensional
pattern space.

Each of our maps has $50 \times 30$ nodes, which gives a useful
trade-off between resolution and speed. For the first phase, the
initial size of the neighbourhood corresponds to the size of the
map. The number of iterations is $1.5 \times 10^6$ and the learning rate
is 10 per cent. The first phase establishes the large-scale ordering
map. For the second phase, the initial size of the neighbourhood is 3,
the number of iteration is $1.5 \times 10^7$ and the learning rate is 5
per cent. The second-phase fine-tunes the ordering on the map.

\section{Cartography}

A SOM constructed using $1.3 \times 10^5$ high quality lightcurves
from the OGLE II dataset is shown in Fig.~\ref{fig:firstsom}.  The $50
\times 30$ nodes of the lattice are the cluster centres. The ordering
is topological -- like maps of the London or Paris metro networks --
and so the distance separating nearby nodes is not faithful. Rather,
the distance is indicated by the colour coding on the upper-left
panel. Neighbouring nodes that are truly close together are coloured
yellow, while those truly far apart are coloured green or black. So, a
grouping of yellow nodes on the map implies a tight clustering of
nodes in the pattern space. Each pattern is mapped onto a node
associated with the nearest weight vector (that is, the one with the
smallest Euclidean distance from the pattern). This distance is
referred to as quantization error. An alternative way of viewing the
same map is shown in Fig.~\ref{fig:secondsom}, in which the nodes are
colour-coded according to the number of hits (or mapped
lightcurves). Blue denotes a node onto which many lightcurves are
mapped and so is a clustering centre. Black denotes empty and dark-red
very sparsely sampled nodes.

\begin{figure*}[!ht]
  \begin{center}
\vspace{0.25\hsize}\centering{\tt figure6.gif}\vspace{0.25\hsize}
  \end{center}
  \caption{The nodes onto which the known small-amplitude
red giant variable stars of type B are mapped are marked with a number.
The convention for the colours of the numbers is given by the key
in the upper right panel.}
  \label{fig:sixthsom}
\end{figure*}

Fig.~\ref{fig:thirdsom} shows the same SOM with again the nodes
colour-coded according to distance (as in Fig.~\ref{fig:firstsom}).
Overplotted are contours of the bump statistic, with blue denoting a
dip and yellow a bump. Also shown on the SOM are the locations of 2580
known eclipsing binaries found by Wyrzykowski et al. (2003) in OGLE
data towards the Large Magellanic Cloud (and so distinct from the
dataset towards the bulge). If an eclipsing binary is mapped to a
node, then the node carries a number. Shades of grey correspond to the
number of eclipsing binaries (white being many and being few).  The
number on the node is the percentile (from 1 to 100) of the node
quantization error distribution corresponding to the median error of
the eclipsing binaries. In other words, a `50' or smaller number means
the eclipsing binary lightcurve looks very similar to the majority of
the lightcurves mapped onto a node, while a `100' means it looks
rather different. So, for example, large numbers of the LMC eclipsing
binaries lightcurves are concentrated on nodes with map coordinates
(13,10) or (14,9) or (16,12). Returning to the bulge dataset, sample
lightcurves mapped onto (14,9) are shown in Fig.~\ref{fig:lcs} -- as
expected, they are all eclipsing binaries with comparable lightcurve
shapes.  Reassuringly, they coincide closely with the blue contours of
small bump statistic.  The SOM has successfully clustered similar
lightcurves and mapped them onto nearby nodes.

\begin{figure}[!ht]
  \begin{center}
\vspace{0.5\hsize}\centering{\tt figure7.gif}\vspace{0.5\hsize}
  \end{center}
  \caption{Sample lightcurves mapped onto the node (14, 9). They are
almost all eclipsing binaries.} 
\label{fig:lcs}
\end{figure}

Fig.~\ref{fig:fourthsom} shows the same underlying SOM with contours
of the bump statistic overplotted. Suppose we wish to identify the
nodes corresponding to the classical and dwarf novae.  Strongly peaked
lightcurves have large values of the bump statistic and so are
enclosed by yellow contours.  There are 32 eruptive cataclysmic
variables already identified toward the Galactic Bulge by Cieslinki et
al. (2003). As expected, we see that they are indeed mapped to the
nodes enclosed by the yellow contours.

Fig.~\ref{fig:fifthsom} shows the same underlying SOM with contours of
the amplitude overplotted. Suppose we wish to identify the nodes
corresponding to the 8970 OGLE small-amplitude red giant (OSARG)
variables in the Galactic bar identified by Wray, Eyer \& Paczy\'nski
(2004). The authors outline two main classes of OSARG variables (type
A and B) according to their amplitude and colour. Then, as the blue
contours in Fig.~\ref{fig:fifthsom} enclose the nodes on which the
small amplitude variables are mapped, this is the place where the
OSARGs of type A are expected. Fig.~\ref{fig:sixthsom} shows the
locations of the known OSARG variables of type B. They occupy the
region located next to the cloud of type A members but prefer the
nodes with larger amplitude. Again, the SOM has recognised similar
patterns and successfully clustered them.

As far as the variable star classification is concerned, the
visualization is an important, albeit intermediate, step in the data
processing. The final goal is to assign variability class membership.
To enable quantitative comparison, we develop the idea of tree
diagrams, as shown in Fig.~\ref{fig:tree}. We start with each node
representing a class.  Then the nodes are clustered by regarding the
nodes within a given distance threshold as identical. As the distance
threshold increases, clusters come together to form super-clusters,
and so on. At each distance threshold, the tree diagram shows the
branching and hence the number of distinct data clusters. In
Fig.~\ref{fig:tree}, the branches are colour-coded according to the
number of objects, with white representing most abundant and black
least. The best choice of distance threshold to identify a class is
given by selecting long branches with constant colour. In
Fig~\ref{fig:tree_eclips}, we show a tree diagram restricted to the
nodes identified with (LMC) eclipsing binaries. We see two major
classes that are easily identifiable from their white/yellow colour,
together with a number of well-defined minor classes.

So, SOMs carry out clustering very quickly. The SOM has enabled us to
identify new candidates for classical and dwarf novae, eclipsing
binaries and small-amplitude red giant variables in the OGLE-II
dataset with very little work. In general, very large datasets can be
processed extremely rapidly with SOMs and the broad features of the
data readily extracted. As we have illustrated here, their ideal role
is to conduct a ``quick look'' through the dataset, identifying the
most prominent features.

\begin{figure*}[!ht]
  \begin{center}
\vspace{0.25\hsize}\centering{\tt figure8.gif}\vspace{0.25\hsize}
  \end{center} \caption{A tree showing the main strands of clustering
  in the SOM shown in Fig. 1. Plotted vertically is the distance
  threshold required for identification as member of a cluster. The
  horizontal scale is arbitrary. The colour of the branch represents
  the logarithm of the number of lightcurves. Long strands with a
  constant colour are well-defined and so astrophysically meaningful
  clusters.}
\label{fig:tree}
\end{figure*}
\begin{figure*}[!ht]
  \begin{center} 
\vspace{0.25\hsize}\centering{\tt figure9.gif}\vspace{0.25\hsize}
\end{center}
\caption{A tree restricted to the nodes corresponding to eclipsing
binaries. There are two strands with white or yellow colour,
indicating two dominant classes of eclipsing binaries. Subsidiary
branches correspond to further sub-dominant classes.}
\label{fig:tree_eclips}
\end{figure*}
\begin{figure*}[!ht]
  \begin{center}
    \leavevmode
\centerline{\epsfig{file=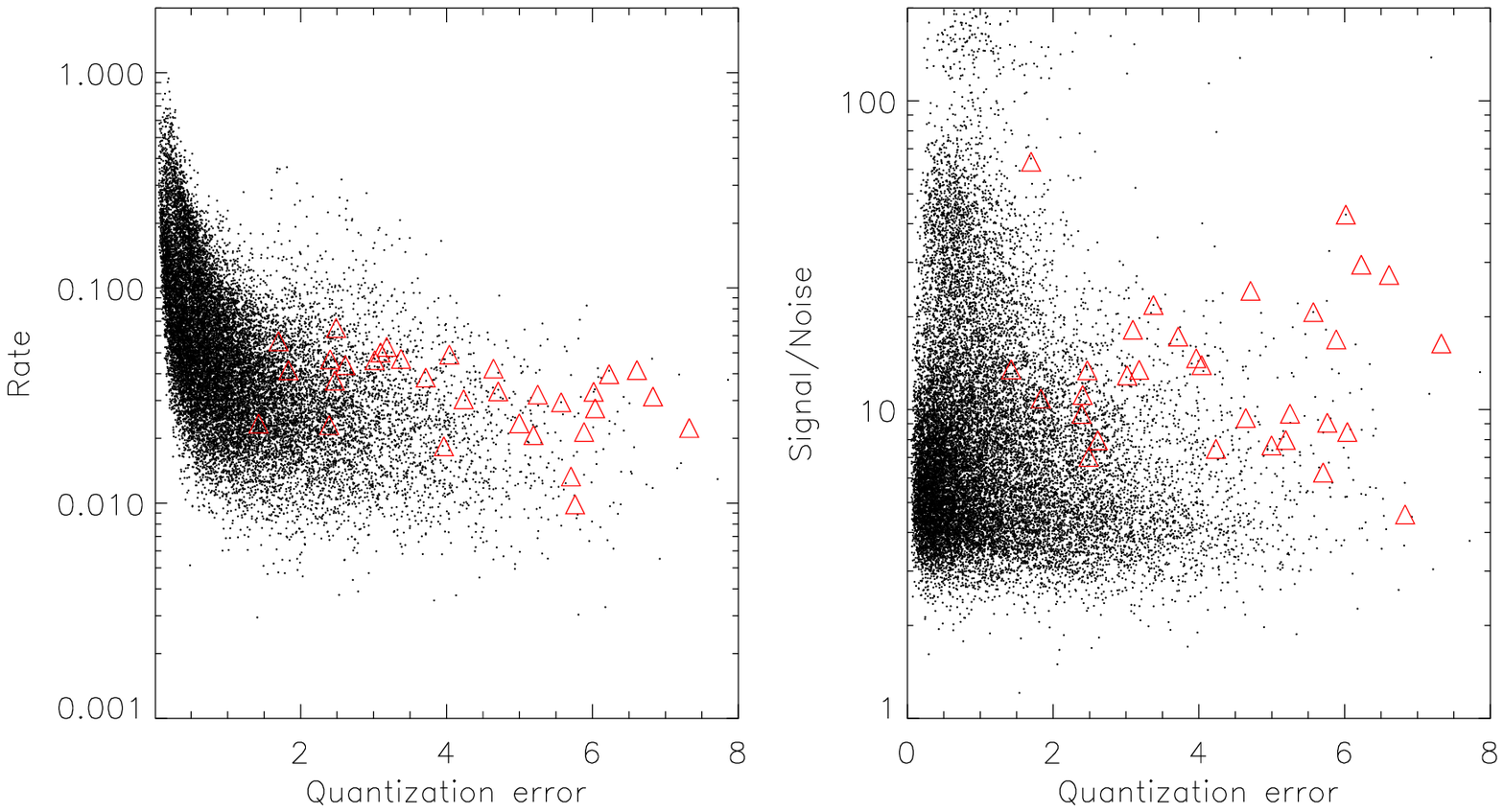,angle=0, width=0.85\textwidth}}
  \end{center} \caption{Left: Lightcurves plotted in the space of
  pattern rate versus quantization error. Note the change of slope at
  a quantization error of $\sim 2$. This marks the threshold error for
  novelty detection.  Right: Lightcurves plotted in the space of
  signal-to-noise ratio (S/N) versus quantization error. The
  conditions $S/N > 10$ and quantization error $> 2$ carry out novelty
  detection. The known OGLE II dwarf and classical novae are marked
  with red triangles. (For reasons of clarity, only a tenth of the
  dataset is plotted). }
\label{fig:alert}
\end{figure*}

\section{Alerts}

How do SOMs single out rare patterns for future study and follow-up
work? How can they perform alerts? Events that are rare may be
scattered over the SOM, as they may not be common enough to warrant
allocation of a node. However, they are recognisable through their
large quantization error. They are different from the most common
patterns mapped to the nodes.

An unusual pattern (such as a supernova-like lightcurve) has an
additional property. Not merely is it distance from the node onto
which it is mapped, it is also distant from all nodes.  For each data
pattern, suppose the 10 closest nodes are found and sorted in the
order of increasing distance. Than a linear fit of the distance
(scaled by the smallest one) versus node number is produced. Let us
call the slope the ``pattern rate''.  A rare pattern is well away from
all the nodes of the lattice and so the slope is very small.

Graphs of quantization error versus pattern rate and quantization
error versus signal-to-noise ratio are shown in Figs.~\ref{fig:alert}.
Suppose the threshold is placed at a quantization error of 2 (this is
the approximate value of the quantization error where the rate changes
slope). Suppose also the threshold is placed at $S/N > 10$.  The
combined cuts reduce the dataset to about one per cent of the
total. These are the most discrepant lightcurves with highest
signal-to-noise ratio.  These are exactly the lightcurves from which
we wish the alerts to come. As an example of this, we show overplotted
on Fig.~\ref{fig:alert} red triangles which correspond to the
locations of the known dwarf and classical novae in the OGLE II
data. (These are a proxy for supernovae lightcurves).

Of course, further experimentation is needed to decide whether such
cuts are enough on their own to issue alerts, or whether such cuts
provide a drastic reduction in the data but further, more
sophisticated processing is required. Even if the latter turns out to
be true, the SOM has played a crucial role in eliminating almost all
the common patterns, allowing us to concentrate on the discrepant few.

What needs to change for application to the GAIA dataset?

First, the input vector needs to be constructed from the GAIA
photometry/spectroscopy/astrometry and may be different for different
objects (supernovae, microlensing and so on).  Experimentation with
different pattern spaces is needed to identify the most telling
combination of observables for diagnostic purposes.  Second, the
lightcurves in our experiments have substantial gaps (6 month periods)
but comprise $\sim 3$ years data. At the beginning of the GAIA
mission, the timeseries is short.  As the GAIA great circle transits
occur, a longer timeseries becomes available.  So, objects follow
tracks in the SOM, settling down to a stable winning node after a
number of transits. Preliminary experimentation suggsts that the
settling-down time is between 6 months and a year.

In other words, a prototype of the GAIA alert system might be as
follows.

\noindent
[1] Every 6 hours, a SOM is built from the GAIA datastream. The
discrepant patterns are extracted with cuts on signal-to-noise and
quantization error. Also extracted are the common patterns
corresponding to known types of variable stars.

\noindent
[2] The discrepant patterns are cross-checked against a catalogue
of known stellar variables. Some of the stellar variables 
can be pre-loaded from existing surveys of variable stars (such
as those available from the microlensing surveys). This catalogue
however will be incomplete at the beginning of the mission and
so will need to be up-dated every 6 hours with new variables
identified by the SOM. 

\noindent
[3] If the discrepant patterns are not in the catalogue of variable
stars, then they are candidates for alerts and need to be looked at
very closely. It may be that there is already enough confidence that
the object needs ground-based follow-up to issue an alert. It may be
that further tests are needed for specific classes of object.

\section{Conclusions}

SOMS are a powerful way to take a quick-look at the data. They provide
a broad brush clustering of the main types of pattern very
quickly. They are ideal for Petabyte datasets (like GAIA). This is
because they are fast, unsupervised and make no prior assumptions
about the data. They represent the diametrically opposed viewpoint to
Bayesian methods, in which the data are analysed exploiting any prior
knowledge.

As an illustration of the technique, the OGLE II difference image
lightcurves of sources towards the bulge have been analysed and the
main classes of stellar variability identified. This gives new
candidates for classical and dwarf novae, eclipsing binaries and
small-amplitude red giant variables. As a way of identifying the
most robust data clusters, we introduced the idea of tree diagrams,
which trace the clustering as a function of distance threshold.

We have argued that SOMS are a powerful way of implementing novelty
detection as well. The number of nodes is roughly the number of
distinct classes and is always much smaller than the number of
different patterns.  The nodes represent the most abundant
patterns. If a pattern is rare, then necessarily there will be no node
allocated near to it. So, such patterns are identifiable through their
large ``quantization error''. Cuts on the quantization error and the
signal-to-noise ratio can substantially reduce the amount of
data. This may already be enough to identify the high-quality
discrepant patterns on which we wish to alert.

\section*{Acknowledgments}
This research was supported by the Particle Physics and Astronomy
Research Council of the United Kingdom. We acknowledge useful
conversations with Laurent Eyer and Piotr Popowski.

\end{document}